\documentclass[12pt]{article}

\usepackage[utf8]{inputenc}
\usepackage{cancel}
\usepackage{amssymb}
\usepackage{slashed}
\usepackage{soul}
\usepackage{tabularx}
\usepackage{xcolor}
\usepackage{booktabs}
\usepackage{subcaption} 
\usepackage{colortbl}
\usepackage{authblk}
\DeclareUnicodeCharacter{00A0}{ }

\newcolumntype{C}{>{\centering\arraybackslash}X}
\usepackage[T1]{fontenc}
\usepackage{epsfig}
\usepackage{latexsym}
\usepackage{graphicx}
\usepackage{amsmath}
\usepackage{amsfonts}   
\usepackage{amssymb}    
\usepackage{float}
\usepackage{bm}
\usepackage{url}
\usepackage{hyperref} 
\usepackage[nodisplayskipstretch]{setspace}
\def\order#1{\ensuremath{{\cal O}(#1)}}
\def\lsim{\raise0.3ex\hbox{$\;<$\kern-0.75em\raise-1.1ex\hbox{$\sim\;$}}}
\def\gsim{\raise0.3ex\hbox{$\;>$\kern-0.75em\raise-1.1ex\hbox{$\sim\;$}}}

\def    \beq            {\begin{equation}}
\def    \eeq            {\end{equation}}
\def    \bea           {\begin{eqnarray}}
\def    \eea           {\end{eqnarray}}

\def \mn{\mu\nu{\rm SSM}}

\def\g2{{\rm GeV}^2}

\def\sw2{sin^2 \theta_w}

\def\a^tau{\alpha_{\tau}}

\def\beq{\begin{equation}}
\def\eeq{\end{equation}}
\def\beqa{\begin{eqnarray}}
\def\eeqa{\end{eqnarray}}

\newcommand{\tev}{\,\textrm{TeV}}
\newcommand{\gev}{\,\textrm{GeV}}

\newcommand{\newc}{\newcommand}
\newc\BR{BR}
\newc{\akappa}{A_{\kappa} }
\newc\deltagmtwo{\delta (g-2)_{\mu}} 
\newc\deltaamu{\Delta a_{\mu}}

\def\anti{\overline}

\def\la{\lambda}

\def\ka{\kappa}

\newc{\haa}{BR\(h_1\to a_1 a_1\)}
\newc{\abb}{BR\(a_1\to b\anti{b}\)}
\newc{\hbb}{BR\(h_1\to b\anti{b}\)}
\newc{\abund}{\Omega h^2}
\newc\bsgamma{b\rightarrow s \gamma }
\newc\bxsgamma{\overline{B}\rightarrow X_{s}\gamma}
\newc\brbsgamma{\BR(\overline{B}\rightarrow X_s\gamma)}

\newc{\Fermi}{\textit{Fermi}-}

\allowdisplaybreaks

\usepackage{feynmp}
\graphicspath{{./Figures/Diagrams/}}
\usepackage{enumitem}

\setul{0.5ex}{0.3ex}
\definecolor{Blue}{rgb}{0,0.0,1}
\setulcolor{Blue}

\usepackage{array, makecell}
\usepackage{boldline}
\usepackage[nosort]{cite}

\title{{\bf Searching for sbottom LSP at the LHC
}}

\author[a]{Paulina~Knees\thanks{pknees@df.uba.ar}}
\author[b,c,d]{Essodjolo Kpatcha\thanks{kpatcha.essodjolo@uam.es}}
\author[e]{I\~naki Lara\thanks{inaki.lara@fuw.edu.pl}}
\author[a,f]{Daniel~E.~L\'opez-Fogliani\thanks{daniel.lopez@df.uba.ar}}
\author[b,c]{Carlos~Mu\~noz\thanks{c.munoz@uam.es}}

 \affil[a]{Instituto de Física de Buenos Aires UBA \& CONICET, Departamento de Física, Facultad de Ciencia Exactas y Naturales, Universidad de Buenos Aires, 
 1428 Buenos Aires, Argentina}
 \affil[b]{Departamento de F\'{\i}sica Te\'{o}rica, Universidad Aut\'{o}noma de Madrid (UAM),
Campus~de~Cantoblanco, 28049 Madrid, Spain}
  \affil[c]{Instituto de F\'{\i}sica Te\'{o}rica (IFT) UAM-CSIC, Campus de Cantoblanco, 28049 Madrid, Spain}
  \affil[d]{Université Paris-Saclay, CNRS/IN2P3, IJCLab, 91405 Orsay, France}
 \affil[e] 
  {Faculty of Physics, University of Warsaw, Pasteura 5, 02-093 Warsaw, Poland}
 \affil[f]{
 {Pontificia Universidad Católica Argentina, 
 Av. Alicia Moreau de Justo~1500, 
 1107~Buenos~Aires, Argentina}}

\date{}

\begin{document}
\maketitle
\begin{abstract}

Assuming that the sbottom is the lightest supersymmetric particle (LSP), we carry out
an analysis of the relevant signals expected at the LHC.
The discussion is established in the framework of the $\mu\nu$SSM, where the presence of $R$-parity violating couplings involving right-handed neutrinos solves simultaneously the $\mu$-problem and the accommodation of neutrino masses and mixing angles.
The sbottoms are pair produced at the LHC, decaying to a lepton and a top quark or a neutrino and a bottom quark. 
The decays can be prompt or displaced, depending on the regions of the parameter space of the model. 
We focus the analysis on the right sbottom LSP, since the left sbottom is typically heavier than the left stop because of the D-term contribution.
We compare the predictions of this scenario with ATLAS and CMS searches for prompt and long-lived particles.
To analyze the parameter space
we sample the $\mu\nu$SSM for a right sbottom LSP,
paying special attention to reproduce the current experimental data on neutrino and Higgs physics, as well as flavor observables.
For displaced (prompt) decays, our results translate into lower limits on the mass  of the right sbottom LSP of about $1041$~GeV ($1070$~GeV). 
The largest possible value found for the decay length is about $3.5$~mm.

\end{abstract}


\newpage 

\tableofcontents 


\section{Introduction}

The `$\mu$~from~$\nu$' Supersymmetric Standard Model
($\mn$)~\cite{LopezFogliani:2005yw,Escudero:2008jg} (for a recent review, see Ref.~\cite{Lopez-Fogliani:2020gzo}) is a predictive model alternative to the 
Minimal Supersymmetric Standard Model (MSSM)~\cite{Nilles:1983ge,Barbieri:1987xf,Haber:1984rc,Gunion:1984yn, Martin:1997ns} 
and the Next-to-MSSM (NMSSM)~\cite{Maniatis:2009re,Ellwanger:2009dp}.
It solves the
$\mu$-problem and the $\nu$-problem (neutrino masses) simultaneously, without
the need to introduce additional energy scales beyond the supersymmetry (SUSY)-breaking scale. In contrast to the MSSM, and the NMSSM, $R$-parity
and lepton number are not conserved,
leading to a completely different
phenomenology characterized by distinct prompt or displaced
decays of the lightest supersymmetric particle (LSP),
producing multi-leptons/jets/photons with small/moderate missing transverse energy (MET) from 
neutrinos~\cite{Ghosh:2017yeh,Lara:2018rwv,Lara:2018zvf,Kpatcha:2019gmq,Kpatcha:2019pve,Heinemeyer:2021opc,Kpatcha:2021nap}.
The smallness of neutrino masses is directly related with the low decay width of the LSP. Actually, it is also related to the existence of possible candidates for decaying dark matter in the model.
This is the case of 
the gravitino~\cite{Choi:2009ng,GomezVargas:2011ph,Albert:2014hwa,GomezVargas:2017,Gomez-Vargas:2019mqk}, or the axino~\cite{Gomez-Vargas:2019vci}, with lifetimes greater than the age of the Universe. 
It is also worth mentioning concerning cosmology, that baryon asymmetry might be realized in the
$\mn$ through electroweak (EW) baryogenesis~\cite{Chung:2010cd}.  
The EW sector of the $\mn$ can also explain~\cite{Kpatcha:2019pve,Heinemeyer:2021opc}
 the longstanding
discrepancy between the experimental result for the anomalous magnetic
moment of the muon~\cite{Abi:2021gix,Albahri:2021ixb} and its SM prediction~\cite{Aoyama:2020ynm}.\footnote{
In this work we will not try to explain it
since we are interested in the analysis of a sbottom LSP through the decoupling of the rest of the SUSY spectrum.}

Because of $R$-parity violation (RPV) in the $\mn$, basically all SUSY particles are candidates for the LSP, and therefore analyses of the LHC phenomenology associated to each candidate are necessary to test them.
This crucial task, given the current experimental results on SUSY searches, has been mainly concentrated on the EW sector of the $\mn$, analyzing left sneutrinos, the right smuon and the bino as candidates for the LSP~\cite{Ghosh:2017yeh,Lara:2018rwv,Lara:2018zvf,Kpatcha:2019gmq,Kpatcha:2019pve,Heinemeyer:2021opc}.
More recently, the color sector of the $\mn$ has started to be analyzed.
In particular, in Ref.~\cite{Kpatcha:2021nap}
the SUSY partners of the top quark as LSP candidates, i.e. the left and right stops, were considered. 
 The aim of this work is to continue with the systematic analysis of the color sector of the $\mn$, focusing now on the right sbottom as the LSP.
 As we will discuss, 
although the left sbottom can also be light, the D-term contribution makes the left stop lighter.

Thus, we will study the constraints on the parameter space of the model by sampling it to get the right sbottom as the LSP in a wide range of masses. We will pay special attention
to reproduce 
{neutrino masses and mixing angles~\cite{Capozzi:2017ipn,deSalas:2017kay,deSalas:2018bym,Esteban:2018azc,deSalas:2020pgw,Esteban:2020cvm}.} 
In addition, we will impose on the resulting parameters agreement with Higgs data as well as with flavor observables.

The paper is organized as follows. In Section~\ref{sec:model}, we will review the $\mn$ and its relevant parameters for our analysis of neutrino, neutral Higgs and sbottom sectors.
In Section~\ref{sec:sbottoms},
we will introduce the phenomenology of the sbottom LSP, studying its pair production channels at the LHC and its signals. The latter consist of displaced vertices with 
a lepton and a top quark or a neutrino and a bottom quark.
In Section~\ref{strategy}, we will discuss the strategy that we will employ to
perform scans searching for points of the parameter space of our scenario compatible with current experimental data on neutrino and Higgs physics, as well as flavor observables such as 
$B$ and $\mu$ decays.
The results of these scans will be presented 
in Section~\ref{sec:results}, and applied to show the 
current reach of the LHC search on the parameter space of the sbottom LSP based {on ATLAS and CMS 
results~\cite{CMS:2019qjk,ATLAS:2017tny,ATLAS:2021yij,ATLAS:2013lcn}.}
Finally, 
our conclusions are left for Section~\ref{sec:conclusions}.

\section{The $\mu\nu$SSM}
\label{sec:model}

In the $\mn$~\cite{LopezFogliani:2005yw,Escudero:2008jg,Lopez-Fogliani:2020gzo}, the particle content of the MSSM
is extended by RH neutrino superfields $\hat \nu^c_i$. 
{The simplest superpotential of the model 
is the following~\cite{LopezFogliani:2005yw,Escudero:2008jg,Ghosh:2017yeh}: 
\bea
W &=&
\epsilon_{ab} \left(
Y_{e_{ij}} 
\, \hat H_d^a\, \hat L^b_i \, \hat e_j^c +
Y_{d_{ij}} 
\, 
\hat H_d^a\, \hat Q^{b}_{i} \, \hat d_{j}^{c} 
+
Y_{u_{ij}}  
\,  
\hat H_u^b\, \hat Q^{a} 
\, \hat u_{j}^{c}
\right)
\nonumber\\
&+& 
\epsilon_{ab} \left(
Y_{{\nu}_{ij}} 
\, \hat H_u^b\, \hat L^a_i \, \hat \nu^c_j
-
\lambda_i \, \hat \nu^c_i\, \hat H_u^b \hat H_d^a
\right)
+
\frac{1}{3}
\kappa_{ijk}
\hat \nu^c_i\hat \nu^c_j\hat \nu^c_k,
\label{superpotential}
\eea
where the summation convention is implied on repeated indices, with $i,j,k=1,2,3$ the usual family indices of the SM 
and $a,b=1,2$ $SU(2)_L$ indices with $\epsilon_{ab}$ the totally antisymmetric tensor, $\epsilon_{12}= 1$. 
}

{Working in the framework of a typical low-energy SUSY, the Lagrangian  containing the soft SUSY-breaking terms related to $W$ 
is given by:
\bea
-\mathcal{L}_{\text{soft}}  =&&
\epsilon_{ab} \left(
T_{e_{ij}} \, H_d^a  \, \widetilde L^b_{iL}  \, \widetilde e_{jR}^* +
T_{d_{ij}} \, H_d^a\,   \widetilde Q^b_{iL} \, \widetilde d_{jR}^{*} 
+
T_{u_{ij}} \,  H_u^b \widetilde Q^a_{iL} \widetilde u_{jR}^*
+ \text{h.c.}
\right)
\nonumber \\
&+&
\epsilon_{ab} \left(
T_{{\nu}_{ij}} \, H_u^b \, \widetilde L^a_{iL} \widetilde \nu_{jR}^* 
- 
T_{{\lambda}_{i}} \, \widetilde \nu_{iR}^*
\, H_d^a  H_u^b
+ \frac{1}{3} T_{{\kappa}_{ijk}} \, \widetilde \nu_{iR}^*
\widetilde \nu_{jR}^*
\widetilde \nu_{kR}^*
\
+ \text{h.c.}\right)
\nonumber\\
&+&   
m_{\widetilde{Q}_{ijL}}^2
\widetilde{Q}_{iL}^{a*}
\widetilde{Q}^a_{jL}
{+}
m_{\widetilde{u}_{ijR}}^{2}
\widetilde{u}_{iR}^*
\widetilde u_{jR}
+ 
m_{\widetilde{d}_{ijR}}^2 
\widetilde{d}_{iR}^* 
\widetilde d_{jR}
+
m_{\widetilde{L}_{ijL}}^2 
\widetilde{L}_{iL}^{a*}  
\widetilde{L}^a_{jL}
\nonumber\\
&+&
m_{\widetilde{\nu}_{ijR}}^2
\widetilde{\nu}_{iR}^*
\widetilde\nu_{jR} 
+
m_{\widetilde{e}_{ijR}}^2
\widetilde{e}_{iR}^*
\widetilde e_{jR}
+ 
m_{H_d}^2 {H^a_d}^*
H^a_d + m_{H_u}^2 {H^a_u}^*
H^a_u
\nonumber \\
&+&  \frac{1}{2}\, \left(M_3\, {\widetilde g}\, {\widetilde g}
+
M_2\, {\widetilde{W}}\, {\widetilde{W}}
+M_1\, {\widetilde B}^0 \, {\widetilde B}^0 + \text{h.c.} \right).
\label{2:Vsoft}
\eea

In the early universe not only the EW symmetry is broken, but in addition to the neutral components of the Higgs doublet fields $H_d$ and $H_u$ also the left and right sneutrinos $\widetilde\nu_{iL}$ and $\widetilde\nu_{iR}$
acquire a vacuum expectation value (VEV). {With the choice of CP conservation, they develop real VEVs denoted by:  
\begin{eqnarray}
\langle H_{d}^0\rangle = \frac{v_{d}}{\sqrt 2},\quad 
\langle H_{u}^0\rangle = \frac{v_{u}}{\sqrt 2},\quad 
\langle \widetilde \nu_{iR}\rangle = \frac{v_{iR}}{\sqrt 2},\quad 
\langle \widetilde \nu_{iL}\rangle = \frac{v_{iL}}{\sqrt 2}.
\end{eqnarray}
The EW symmetry breaking is induced by the soft SUSY-breaking terms
producing
$v_{iR}\sim {\order{1 \tev}}$ as a consequence of the right sneutrino minimization equations in the scalar potential~\cite{LopezFogliani:2005yw,Escudero:2008jg,Ghosh:2017yeh}.
Since $\widetilde\nu_{iR}$ are gauge-singlet fields,
the $\mu$-problem can be solved in total analogy to the
NMSSM
through the presence in the superpotential (\ref{superpotential}) of the trilinear 
terms $\lambda_{i} \, \hat \nu^c_i\,\hat H_u \hat H_d$.
Then, the value of the effective $\mu$-parameter is given by 
$\mu=
\la_i v_{iR}/\sqrt 2$.
These trilinear terms also relate the origin of the $\mu$-term to the origin of neutrino masses and mixing angles, since neutrino Yukawa couplings 
$Y_{{\nu}_{ij}} \hat H_u\, \hat L_i \, \hat \nu^c_j$
 are present in the superpotential {generating Dirac masses for neutrinos, 
$m_{{\mathcal{D}_{ij}}}\equiv Y_{{\nu}_{ij}} {v_u}/{\sqrt 2}$.}
Remarkably, in the $\mu\nu$SSM it is possible to accommodate neutrino masses
and mixings in agreement with experiments~\cite{Capozzi:2017ipn,deSalas:2017kay,deSalas:2018bym,Esteban:2018azc} via an EW seesaw
mechanism dynamically generated during the EW symmetry breaking~\cite{LopezFogliani:2005yw,Escudero:2008jg,Ghosh:2008yh,Bartl:2009an,Fidalgo:2009dm,Ghosh:2010zi,Liebler:2011tp}. The latter takes place 
through the couplings
$\kappa{_{ijk}} \hat \nu^c_i\hat \nu^c_j\hat \nu^c_k$,
giving rise to effective Majorana masses for RH neutrinos
${\mathcal M}_{ij}
= {2}\kappa_{ijk} {v_{kR}}/{\sqrt 2}$.
Actually, this is possible at tree level even with diagonal Yukawa couplings~\cite{Ghosh:2008yh,Fidalgo:2009dm}.
It is worth noticing here that the neutrino Yukawas discussed above also generate the effective bilinear terms $\mu_i \hat H_u\, \hat L_i $
with $\mu_i=Y_{{\nu}_{ij}} {v_{jR}}/{\sqrt 2}$,
used in the bilinear RPV model (BRPV)~\cite{Barbier:2004ez}.

We conclude therefore, that the $\mn$ solves not only the
$\mu$-problem, but also the $\nu$-problem, without
the need to introduce energy scales beyond the SUSY-breaking one.

The parameter space of the $\mn$, and in particular the
neutrino, neutral Higgs and sbottom sectors are 
relevant for our analysis in order to reproduce neutrino and Higgs data, and to obtain in the spectrum a sbottom as the LSP.
In particular, neutrino and Higgs sectors were discussed in Refs.~\cite{Kpatcha:2019gmq,Kpatcha:2019qsz,Kpatcha:2019pve,Heinemeyer:2021opc}, and we refer the reader to those works for details, although we will summarize the results below. First, we discuss here several simplifications that are convenient to take into account given the large number of parameters of the model.
{Using diagonal mass matrices for the scalar fermions, in order to avoid the
strong upper bounds upon the intergenerational scalar mixing (see e.g. Ref.~\cite{Gabbiani:1996hi}), from the eight minimization conditions with respect to $v_d$, $v_u$,
$v_{iR}$ and $v_{iL}$ to facilitate the computation we prefer to eliminate
the
soft masses $m_{H_{d}}^{2}$, $m_{H_{u}}^{2}$,  
$m_{\widetilde{\nu}_{iR}}^2$ and
$m_{\widetilde{L}_{iL}}^2$
in favor
of the VEVs.
Also, we assume} for simplicity in what follows the flavour-independent couplings and VEVs $\lambda_i = \lambda$,
$\kappa_{ijk}=\kappa \delta_{ij}\delta_{jk}$, and $v_{iR}= v_{R}$. Then, the higgsino mass parameter $\mu$, bilinear couplings $\mu_i$ and Dirac and Majorana masses discussed above are given by:
\bea
\mu=3\la \frac{v_{R}}{\sqrt 2}, \;\;\;\;
\mu_i=Y_{{\nu}_{i}}  \frac{v_{R}}{\sqrt 2}, \;\;\;\;
m_{{\mathcal{D}_i}}= Y_{{\nu}_{i}} 
\frac{v_u}{\sqrt 2}, \;\;\;\;
{\mathcal M}
={2}\kappa \frac{v_{R}}{\sqrt 2},
\label{mu2}    
\eea
where 
we have already used the possibility of having diagonal neutrino Yukawa couplings $Y_{{\nu}_{ij}}=Y_{{\nu}_{i}}\delta_{ij}$ in the $\mn$ in order to reproduce neutrino physics.

\subsection{The neutrino sector}
\label{neutrino}

For light neutrinos, under the above assumptions, one can obtain
the following simplified formula for the effective mass matrix~\cite{Fidalgo:2009dm}:
\begin{eqnarray}
\label{Limit no mixing Higgsinos gauginos}
(m_{\nu})_{ij} 
\approx
\frac{m_{{\mathcal{D}_i}} m_{{\mathcal{D}_j}} }
{3{\mathcal{M}}}
                   \left(1-3 \delta_{ij}\right)
                   -\frac{v_{iL}v_{jL}}
                   {4M}, \;\;\;\;\;\;\;\;
        \frac{1}{M} \equiv \frac{g'^2}{M_1} + \frac{g^2}{M_2},         
\label{neutrinoph2}
  \end{eqnarray}     
where $g'$, $g$ are the EW gauge couplings, and $M_1$, $M_2$ the bino and wino soft {SUSY-breaking masses}, respectively.
This expression arises from the generalized EW seesaw of the $\mn$, where due to RPV the neutral fermions have the flavor composition
$(\nu_{iL},\widetilde B^0,\widetilde W^0,\widetilde H_{d}^0,\widetilde H_{u}^0,\nu_{iR})$.
The first two terms in Eq.~(\ref{neutrinoph2})
are generated through the mixing 
of $\nu_{iL}$ with 
$\nu_{iR}$-Higgsinos, and the third one 
also include the mixing with the gauginos.
These are the so-called $\nu_{R}$-Higgsino seesaw and gaugino seesaw, respectively~\cite{Fidalgo:2009dm}.
One can see from this equation that {once ${\mathcal M}$ is fixed, as will be done in the parameter analysis of Section~\ref{sec:parameter},
the most crucial independent parameters determining {neutrino physics} are}:
\bea
Y_{\nu_i}, \, v_{iL}, \, M_1, \, M_2.
\label{freeparameters}
\eea
Note that this EW scale seesaw implies $Y_{\nu_i}\lsim 10^{-6}$
driving $v_{iL}$ to small values because of the proportional contributions to
$Y_{\nu_i}$ appearing in their minimization equations. {A rough} estimation gives
$v_{iL}\lsim m_{{\mathcal{D}_i}}\lsim 10^{-4}$.

{Considering the normal ordering for the neutrino mass spectrum,
and taking advantage of the 
dominance of the gaugino seesaw for some of the three neutrino families, three
representative type of solutions for neutrino physics using diagonal neutrino Yukawas were obtained in 
Ref.~\cite{Kpatcha:2019gmq}.
In our analysis we will use the so-called type 2 solutions, which have the structure
\bea
M>0, \, \text{with}\,  Y_{\nu_3} < Y_{\nu_1} < Y_{\nu_2}, \, \text{and} \, v_{1L}<v_{2L}\sim v_{3L},
\label{neutrinomassess}
\eea
In this case of type 2, it is easy to find solutions with the gaugino seesaw as the dominant one for the third family. Then, $v_{3L}$ determines the corresponding neutrino mass and $Y_{\nu_3}$ can be small.
On the other hand, the normal ordering for neutrinos determines that the first family dominates the lightest mass eigenstate implying that $Y_{\nu_{1}}< Y_{\nu_{2}}$ and $v_{1L} < v_{2L},v_{3L}$, {with both $\nu_{R}$-Higgsino and gaugino seesaws contributing significantly to the masses of the first and second family}. Taking also into account that the composition of the second and third families in the second mass eigenstate is similar, we expect $v_{2L} \sim v_{3L}$. 
In Ref.~\cite{Kpatcha:2019gmq}, a quantitative analysis of the neutrino sector was carried out, with the result that the hierarchy qualitatively discussed above for Yukawas and VEVs works properly. 
See in particular Fig.~4 of Ref.~\cite{Kpatcha:2019gmq}, where
$\delta m^2=m^2_2-m^2_1$ 
versus $Y_{\nu_{i}}$ and $v_{iL}$ is shown for the scans carried out in that work, using
the results for normal ordering from Ref.~\cite{Esteban:2020cvm}. 

{We will argue in Section~\ref{sec:results} that the 
other two type of solutions of normal ordering for neutrino physics are not going to modify our results.}
The same conclusion is obtained in the case of working with the inverted 
ordering for the neutrino mass spectrum. The structure of the solutions is more involved for this case, because the two heaviest eigenstates are close in mass and the lightest of them has a dominant contribution from the first family. Thus, to choose $Y_{\nu_1}$ as the largest of the neutrino Yukawas helps to satisfy these relations. For the second and third family, a delicate balance between the contributions of $\nu_{R}$-higgsino and gaugino seesaws is needed in order to obtain the correct mixing angles. In particular, a representative type of solutions for the case of inverted ordering has the structure $M>0$, with
$Y_{\nu_3} \sim Y_{\nu_2} < Y_{\nu_1}$, and $v_{1L}<v_{2L}\sim v_{3L}$.

\subsection{The Higgs sector}
\label{sec:higgs}

The neutral Higgses are mixed with right and left sneutrinos, since
the neutral scalars and pseudoscalars in the $\mn$ have the flavor composition
$(H_{d}^0, H_{u}^0, \widetilde\nu_{iR}, \widetilde\nu_{iL}) $.
Nevertheless, the left sneutrinos are basically decoupled from the other states, since
the off-diagonal terms of the mass matrix are suppressed by the small $Y_{\nu_{ij}}$ and $v_{iL}$.
Unlike the latter states, the other neutral scalars can be substantially mixed.
Neglecting this mixing between 
the doublet-like Higgses and the three right sneutrinos, the expression of the tree-level mass of the { SM-like Higgs} is~\cite{Escudero:2008jg}:
\begin{eqnarray}
m_h^2 \approx 
m^2_Z \left(\cos^2 2\beta + 10.9\
{\lambda}^2 \sin^2 2\beta\right),
\end{eqnarray}
where $\tan\beta= v_u/v_d$, and $m_Z$ denotes the mass of the $Z$~boson.
Effects lowering (raising) this mass appear when the SM-like Higgs mixes with heavier (lighter) right sneutrinos. The one-loop corrections are basically determined by 
the third-generation soft {SUSY-breaking} parameters $m_{\widetilde u_{3R}}$, $m_{\widetilde Q_{3L}}$ and $T_{u_3}$
(where we have assumed for simplicity that for all soft trilinear parameters
$T_{ij}=T_{i}\delta_{ij}$).
These three parameters together with the coupling $\lambda$ and $\tan\beta$, are the crucial ones for Higgs physics. 
Their values can ensure that the model contains a scalar boson with a mass around $\sim 125 \gev$ and properties similar to the ones of the SM Higgs boson~\cite{Biekotter:2017xmf,Biekotter:2019gtq,Kpatcha:2019qsz,Biekotter:2020ehh}.

In addition, $\ka$, $v_R$ and the trilinear parameter $T_{\kappa}$ in the 
soft Lagrangian~(\ref{2:Vsoft}),
are the key ingredients to determine
the mass scale of the {right sneutrinos}~\cite{Escudero:2008jg,Ghosh:2008yh}.
For example, for $\lambda\lsim 0.01$ they are basically free from any doublet admixture, and using their minimization equations in the scalar potential
the scalar and pseudoscalar masses can be approximated respectively by~\cite{Ghosh:2014ida,Ghosh:2017yeh}:
\bea
m^2_{\widetilde{\nu}^{\mathcal{R}}_{iR}} \approx   \frac{v_R}{\sqrt 2}
\left(T_{\kappa} + \frac{v_R}{\sqrt 2}\ 4\kappa^2 \right), \quad
m^2_{\widetilde{\nu}^{\mathcal{I}}_{iR}}\approx  - \frac{v_R}{\sqrt 2}\ 3T_{\kappa}.
\label{sps-approx2}
\eea

Finally, $\lambda$ and the trilinear parameter $T_{\lambda}$ 
not only contribute to these masses
for larger values of $\lambda$, but
also control the mixing between the singlet and the doublet states and hence, they contribute in determining their mass scales as discussed in detail in Ref.~\cite{Kpatcha:2019qsz}.
We conclude that the relevant parameters
in the {Higgs (-right sneutrino) sector} are:
\bea
\lambda, \, \kappa, \, \tan\beta, \, v_R, \, T_\kappa, \, T_\lambda, \, T_{u_3}, \,  m_{\widetilde u_{3R}},
\, m_{\widetilde Q_{3L}}.
\label{freeparameterss}
\eea
Note
that the most crucial parameters for the neutrino sector~(\ref{freeparameters}) are basically decoupled from these parameters controlling Higgs physics.
This simplifies the analysis of the parameter space of the model,
as will be discussed in 
Section~\ref{sec:parameter}.

\subsection{The sbottom sector}

\label{sbottom}

The mass matrix 
of the sbottoms includes new terms with respect to the   one of the MSSM~\cite{Escudero:2008jg,Ghosh:2017yeh},
similar to other squarks in the $\mn$. However, these terms are negligible given that they are proportional to the small parameters 
$v_{iL}$. Thus, the sbottom eigenstates of the $\mu\nu$SSM coincide basically with those of the MSSM, and one has the following tree-level mass matrix in the flavor basis ($\widetilde b_L$, $\widetilde b_R$):
\begin{equation}
    m_{\widetilde{b}}^2=\begin{pmatrix}
        m_b^2+m_{\widetilde{Q}_{3L}}^2+\Delta\widetilde{d}_L & m_b X_b \\
        m_b X_b & m_b^2+m_{\widetilde{d}_{3R}}^2+\Delta\widetilde{d}_R
    \end{pmatrix},
\label{eq:matrix-sb}
\end{equation}
where $m_b$ is the bottom-quark mass, $\Delta\widetilde{d}_{L,R}$ denote the D-term contributions 
\bea
\Delta \widetilde d_L = - m^2_Z \left( \frac{1}{2} - \frac{1}{3} \sin^2\theta_W \right) \cos 2\beta, \quad
\Delta \widetilde d_R = - \frac{1}{3}  m^2_Z \sin^2\theta_W \cos 2\beta,
\label{delta-stop}
\eea
with $\theta_W$ the weak-mixing angle,
and $X_b$ the left-right sbottom mixing term
\begin{equation}
    X_b= \frac{T_{d_3}}{Y_{d_3}}- \mu\tan\beta.
\label{eq:Xb}
\end{equation}

As can easily be deduced from Eq.~(\ref{eq:matrix-sb}), the physical sbottom masses are controlled mainly by the value of the soft SUSY-breaking parameters: 
\bea
m_{\widetilde Q_{3L}}, \, m_{\widetilde d_{3R}}, \, T_{d_3}.
\label{freeparametersstop}
\eea
However, the trilinear parameter is typically less relevant than the two mass parameters because it contributes to sbottom masses through the mixing term, which is suppressed by the bottom-quark mass. 
Playing with the values of these parameters, it is straightforward to obtain the lightest eigenvalue dominated either by the left sbottom composition ($\widetilde b_L$) or by the 
right sbottom composition ($\widetilde b_R$). 
Note that in the case of the lightest sbottom mainly $\widetilde b_L$, a small value of the common soft mass $m_{\widetilde Q_{3L}}$ makes $\widetilde t_L$ slightly lighter than 
$\widetilde b_L$ at tree level due to the D-term contribution,
$m_{\widetilde t_L}^{2} = m_{\widetilde b_{L}}^2
+m_W^2 \cos 2\beta$ with $\cos 2\beta <0$.
Thus, in what follows we will focus on the right sbottom LSP, for which a low value of $m_{\widetilde d_{3R}}$ is crucial.

\bigskip

\noindent
In our analysis of 
Section~\ref{sec:results}., 
we will sample the relevant parameter space of the $\mn$, which contains the independent parameters determining neutrino and
Higgs physics in Eqs.~(\ref{freeparameters}) and (\ref{freeparameterss}).
{Nevertheless, the parameters for neutrino physics $Y_{\nu_i}$, $v_{iL}$, $M_1$ and $M_2$ are essentially decoupled from the parameters 
controlling Higgs physics.
Thus, for a suitable choice of the former parameters
reproducing neutrino physics, there is still enough freedom to reproduce in addition Higgs data by playing with 
$\lambda$, $\kappa$, $v_R$, $\tan\beta$, $T_{u_3}$, etc., 
as shown in Refs.~\cite{Kpatcha:2019gmq,Kpatcha:2019pve,Heinemeyer:2021opc}. 
As a consequence, we will not need to scan over most of the latter parameters, relaxing 
our computing task. For this task
{we have employed the 
{\tt Multinest}~\cite{Feroz:2008xx} algorithm as optimizer. {To compute
the spectrum and the observables we have used {\tt SARAH}~\cite{Staub:2013tta} to generate a 
{\tt SPheno}~\cite{Porod:2003um,Porod:2011nf} version for the model.} 
}

\section{Sbottom LSP Phenomenology}
\label{sec:sbottoms}

The production of sbottoms at colliders is dominated by QCD processes, since the RPV contributions to their production are strongly suppressed in the $\mn$. The pair production of colored SUSY particles at large hadron colliders has been extensively studied. Since we do not expect a significant difference from the values predicted in the MSSM, we make use of NNLL-fast-3.0~\cite{Beenakker:2016lwe,Beenakker:1997ut,Beenakker:2010nq,Beenakker:2016gmf} to calculate the number of sbottom pair events produced. In particular,
for our range of interest of sbottom masses between about 200 GeV and 2000 GeV, the production cross section is in the range between 74.4 pb 
and 
 2$\times 10^{-5}$ pb.

\subsection{Decay modes}
\label{subsec:Decaymodes}

 \begin{figure}[t!]
     \centering
     \includegraphics[scale=0.7]{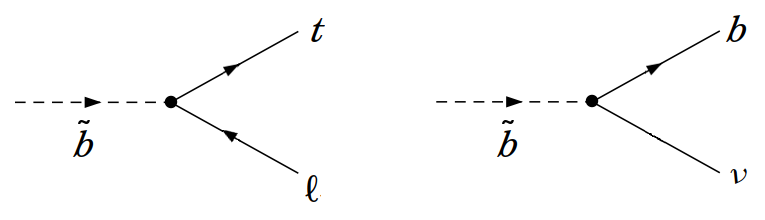}
      \caption{Dominant decay channels in the $\mn$ for a sbottom LSP. (left) Decay to top quark and leptons; (right) Decay to bottom quark and neutrinos. }
     \label{fig:decay-chan}
 \end{figure}

There are two dominant channels for the decay of the right sbottom LSP to standard model particles. Similarly to the stop LSP case~\cite{Kpatcha:2021nap}, the sbottom LSP mainly decays to a quark and a lepton/neutrino. In the case of the decay to quark and leptons, the dominant contribution arises from the top quark, while for the decay to quark and neutrinos it arises from the bottom quark. Both cases are shown in Fig.~\ref{fig:decay-chan}. 

The relevant interactions for our analysis are given in Appendix \ref{appendix}.
There, one can identify the most important contributions for the decays. 
In particular, the relevant diagram shown in Fig.~\ref{fig:decay-chan} left corresponds to the second term multiplying the projector $P_{R}$ in 
Eq.~(\ref{eq:appen-ql}). 
Thus 
it occurs mainly through the Yukawa coupling $Y_{b}$ 
of $\widetilde b$
with $t$ and charged higgsinos,
via the mixing between the latter and $\ell$.
The diagram in
Fig.~\ref{fig:decay-chan} right
 corresponds to the second term multiplying the projector $P_{L}$ (and the
first term multiplying the projector $P_{R}$) in 
Eq.~(\ref{eq:appen-qnu}).
It occurs through 
the 
gauge coupling $g'$ of $\widetilde b$
with $b$ and 
neutral bino (Yukawa coupling $Y_{b}$ 
of $\widetilde b$
with $b$ and neutral higgsinos), via the mixing between bino (higgsinos) and $\nu$.

In the case of (pure) right sbottom LSP,  
the values of the partial decay widths can be approximated, as: 
\begin{equation}
    \Gamma(\widetilde{b}_R \to t \ell_i) \sim \frac{(m_{\widetilde{b}}^2-m_t^2)^2}{16 \pi m_{\widetilde{b}}^3}\left(Y_b \frac{\mu_i}{ \mu}\right)^2,
    \,\,
   \sum_i \Gamma(\widetilde{b}_R \to b \nu_i)\sim \frac{m_{\widetilde{b}}}{16 \pi} \sum_{i}\left[ \left(\frac{\sqrt{2}}{3}g'U_{i4}^V\right)^2+\left(Y_b U_{i6}^V \right)^2\right]
   \label{gamma-bnu}
\end{equation}
As discussed in Appendix~\ref{appendix}, $U^V$ is 
the matrix which diagonalizes the mass matrix for the 
neutral fermions, and the above entries $U^V_{i4}$ and {$U^V_{i6}$}, corresponding to the mixing between neutrinos and
bino and neutrinos and neutral higgsino $\widetilde H^0_d$, respectively,
can 
be approximated as 
\begin{eqnarray}
U^V_{i4}\approx\frac{{-}g'}{M_1}\sum_{l}{\frac{v_{lL}}{\sqrt{2}}U^{\text{\tiny{PMNS}}}_{il}},\quad \quad
{U^V_{i6}\approx\ \frac{1}{\mu}\sum_l \frac{\mu_l}{\sqrt{2}} U^{\text{\tiny{PMNS}}}_{il}},
\label{--sneutrino-decay-width-2nus2}
\end{eqnarray}
where $U^{\text{\tiny{PMNS}}}_{il}$ are the entries of the PMNS matrix, with
$i$ and $l$ neutrino physical and flavor indices, respectively.
We also approximate other entries of the matrices involved in the computation 
 (see Appendix~\ref{appendix}), as follows: $Z_{16}^{D}\approx 1$ (pure right sbottom), {$U_{L, 33}^{d} \approx 1$} and
 $U_{R, 33}^{d} \approx 1$ (pure {LH} and RH bottom quarks),
 $U_{L, 33}^{u} \approx 1$ (pure LH top  quark),
  $U_{R, j 5}^{e}\approx {\mu_i}/{\mu}$.
In addition, we use $m_{\widetilde{b}}\gg m_b, m_{\ell}$.
Let us remark nevertheless that the results of Section~\ref{sec:results}
have been obtained using the full tree-level numerical computation of decay widths implemented in {\tt SPheno},
taking also into account the small contamination between left and right sbottoms,
$|Z^D_{36}|^2$. Loop corrections for sbottom decays are negligible since the dominant ones are two body decays. We have checked it numerically. 

As can be easily deduced from Eq.~(\ref{gamma-bnu}), the decay width of $\widetilde{b}_R$ to leptons is smaller than the one to neutrinos for sbottom masses close to the top mass. 
This is qualitatively different from the case of the stop LSP, where the decay width to leptons is larger that the one to neutrinos for stop masses close to the top mass, as can be seen in Fig.~5 of 
Ref.~\cite{Kpatcha:2021nap}.
In our computation we will use a lower bound for the sbottom mass of 200 GeV.

\subsection{LHC searches}
\label{sec:lhc}

The event topologies originated from the sbottom LSP decaying as described in section~\ref{subsec:Decaymodes} will produce signals at hadron colliders detectable with diverse LHC searches. As it is shown in Fig.~\ref{fig:decay-chan}, the possible decays include: the production of a lepton ($e$, $\mu$ or $\tau$) and a top quark or the production of a neutrino and a bottom quark. Consequently, the production of a pair of sbottoms will lead to events of the form: $\bar{t}t\bar{l}l$, $\bar{b}b\bar{\nu}\nu$ or $\bar{t}b\bar{\nu}l$.
In addition, the decay length of the sbottom LSP ranges from sub-mm scale up to $\sim 30$ mm. Therefore, there are different LHC searches that will have the highest sensitivity for each case. We will classify the signals according to the lifetime scale and apply to each one different searches.

\bigskip

\noindent
{\bf Case i) {Non-prompt jets}}
\vspace{0.2cm}

\noindent 
The timing capabilities of the CMS electromagnetic calorimeter 
 allow to discriminate jets arriving at times significantly larger than the traveling times expected for light hadrons, which are moving at velocities close to the speed of light. This time delay can be associated with two effects: First, the larger indirect path formed by the initial trajectory of a long-lived particle plus the subsequent trajectories of the child particles. Secondly, the slower velocity of the long-lived particle due to the high mass compared to light hadrons. Such analysis is performed by the CMS collaboration in the work~\cite{CMS:2019qjk} in the context of long-lived gluinos decaying to gluons and stable gravitinos, excluding gluinos with masses of $\sim2500$ GeV for lifetimes of $\sim1$ m.

The case where the sbottom LSP decays producing a neutrino and a bottom quark with proper decay lengths above $\sim30$ cm will produce a signal similar to the one analyzed in ~\cite{CMS:2019qjk}. For each point analyzed in this search we compare the 95\% observed upper limit on cross section, corresponding to the signal of two delayed jets for a given parent particle mass and $c\tau$, with the prediction of the signal cross section calculated as $\sigma(pp\to\tilde{b}\tilde{b}^*)\times BR(\tilde{b}\to b \nu)^2$.

\bigskip
\vspace{0.6cm}

\noindent 
{\bf Case ii) {Displaced vertices}}
\vspace{0.2cm}

\noindent 
For shorter lifetimes, one can confront the points of the model with the limits from events
with displaced vertices including jets.
The ATLAS search~\cite{ATLAS:2017tny} targets final states with at least one displaced vertex (DV) with a high reconstructed mass
and a large track multiplicity in events with large missing transverse momentum. The search originally targets long-lived massive particles with lifetimes in the range 1-100mm. Thus, this search can be sensitive to the sbottom LSP when $c\tau$ is in this range. However, the signal topologies analyzed in the search do not match the ones originated from the decays shown in Fig.~\ref{fig:decay-chan}. 

To have a reasonable estimate of the exclusion power of this search, we use a recast version of the analysis within 
CheckMATE-LLP~\cite{Desai:2021jsa}. CheckMATE~\cite{Dercks:2016npn,Kim:2015wza} is a universal tool for the recasting of LHC searches in the context of arbitrary new
physics models. It uses the fast detector simulation framework Delphes~\cite{deFavereau:2013fsa} with customized ATLAS detector card and additional built-in tuning for a more accurate reproduction of experimental efficiencies. The validation of the recasted search is discussed in~\cite{Desai:2021jsa}. 
We generate signal Monte Carlo (MC) samples of sbottom pair production with MadGraph5\_aMC@NLO-v3.4.2~\cite{Alwall:2014hca,Alwall:2007fs,Alwall:2008qv} at leading order (LO). The hard event corresponds to tree-level production of sbottom pairs and includes the emission
of up to two additional partons, 
the NNPDF23LO ~\cite{Ball:2012cx,Buckley:2014ana} PDF set is used. Simulated signal events were passed to 
Pythia-8.306~\cite{Sjostrand:2014zea} for parton showering (PS) and hadronization. Jet matching and merging to parton-shower calculations is accomplished by the MLM algorithm~\cite{Mangano:2006rw}.
Sbottom pair-production nominal cross sections are derived at NNLO+NNLL using  NNLL-fast-3.0~\cite{Beenakker:2016lwe,Beenakker:1997ut,Beenakker:2010nq,Beenakker:2016gmf}.
Finally, we process the events generated trough CheckMATE. The results are used to calculate the efficiency ($\epsilon$) of the search, defined as the number of events predicted in the signal region divided by the total number generated events.

We generate samples for values of the mass equal to $[250, 500,750,1000,1500,2000]$ GeV,  $c\tau$ equal to $[1,3,5,10,15,20,40,60]$ mm and for all of the different combinations of decays shown in Fig.~\ref{fig:decay-chan}, and calculate $\epsilon$ for each case. 
For each point tested in this work, we calculate the $\epsilon$ interpolating from each channel and value of mass and $c\tau$, within the set of $\epsilon$ obtained as described above.

Finally, the point is considered excluded if the total number of events predicted in the signal region of the search\cite{ATLAS:2017tny}, calculated as the sum 
 of $\mathcal{L}\times\sigma(pp\to\tilde{b}\tilde{b}^*)\times BR_{channel}\times \epsilon_{channel}$ over all channels, is greater than the 95\% upper limit on signal events, which correspond to approximately 3 events.

\bigskip

\noindent 
{\bf Case iii) {Prompt and nearly-prompt b-jets}}
\vspace{0.2cm}

\noindent 
If the proper decay length of the sbottom LSP is sufficiently short, the LHC searches designed to look for b-tagged jets originated from the decay of short-lived particles will be sensitive to the sbottom signal.

The ATLAS collaboration has shown, in a reanalysis of a selection of searches targeting RPV and RPC SUSY models~\cite{ATLAS:2018yey}, that the impact of the parent particle lifetime over the distribution of the observables used as discriminants in the searches for b-jets+missing transverse energy (MET)~\cite{ATLAS:2017weo}, such as the number of jets, the missing transverse energy (MET), or the effective mass (meff), is unaffected for values of $c\tau< 3$ mm. Moreover, the b-tagging efficiency is improved for decay lengths of the order of millimeters. 
The same considerations can be made for the ATLAS search for sbottoms in events with b-jets+MET~\cite{ATLAS:2021yij}. However, this search includes an additional restriction with respect to~\cite{ATLAS:2017weo}: Jet candidates are reconstructed from charged-particle tracks matched to the hard-scatter vertex with the requirement $|z_0\sin \theta|<$ 2.0 mm, where $z_0$ is their longitudinal impact parameter. To check the compatibility of the signal of the decay of the sbottom LSP with this requirement, we have generated events corresponding to a pair of 1000 GeV sbottoms decaying to bottom quarks and neutrinos, with different values of $c\tau$. In Fig.~\ref{fig:zsintheta} we show the distribution of $|z_0\sin \theta|$ of the charged tracks at truth level.  For values of $c\tau=1$ mm, more than 90\% of the charged tracks satisfy the requirement. A conservative estimate of the sensitivity of this search to slightly displaced sbottoms can be obtained applying the limits from this search only to points where $c\tau\lesssim1$ mm.\footnote{ A small set of points with masses $\sim830$ GeV and $c\tau\in [1.01-1.03]$ mm cannot be excluded by any of the other strategies defined in this section. For those points, we multiply the result by a factor $\epsilon=1-e^{-
\frac{\sqrt{2}\times \mathrm{1 mm}}{c\tau\beta\gamma}
}$ to account for the restriction of only counting the events were the sbottom LSP decays withing 1mm of the production point.}

 \begin{figure}[t!]
     \centering
     \includegraphics[scale=.6]{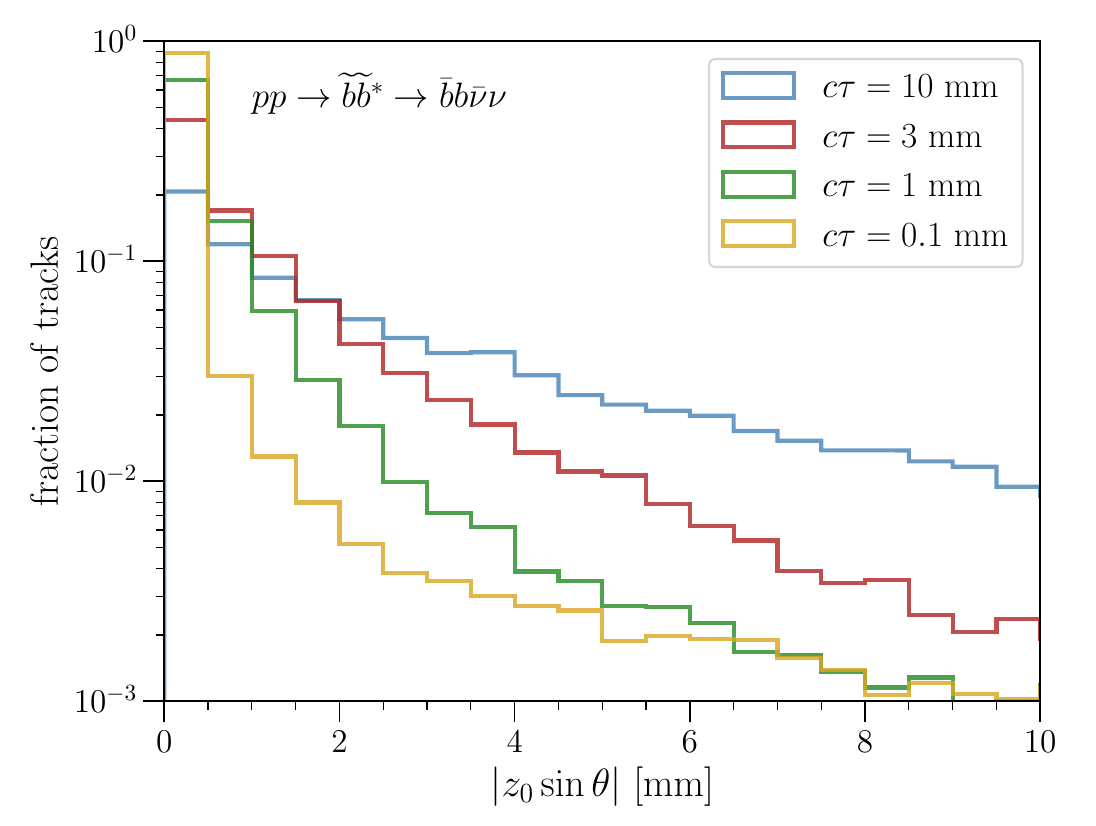}
     \caption{Distribution of the impact parameter of charged tracks for simulated events. }
  \label{fig:zsintheta}
 \end{figure}

We test the points where the sbottom satisfies the requirement on $c\tau$ by comparing the 95\% observed cross-section upper limit, corresponding to sbottom pair production decaying to bottom quarks plus massless neutralinos, in the search~\cite{ATLAS:2021yij} with the prediction of the signal cross section, calculated as $\sigma(pp\to\tilde{b}\tilde{b}^*)\times BR(\tilde{b}\to b \nu)^2$. 

Some of the points explored in this work, with masses between 200 GeV and 400 GeV fall outside of the range of masses analyzed in the ATLAS search~\cite{ATLAS:2021yij}.
An alternative ATLAS search for sbottoms in final states with MET and two b-jets, made with data of accumulated 20.1 $\mathrm{fb}^{-1}$ of pp collisions at $\sqrt{s}=8$ TeV~\cite{ATLAS:2013lcn}, is sensitive to masses between 100 GeV and 800 GeV. Thus, complementing the previous constraints over higher sbottom masses. 
We assume that the same considerations made about the impact of the parent particle lifetime over the kinematic observables based on hadronic activity still hold for this ATLAS analysis. There is an additional requirement, that all jets with $p_T<50\ \mathrm{GeV}$ and $|\eta|<2.5$ are required to have at least one track identified as coming from the primary vertex, otherwise the event is discarded. We expect this requirement to not be sensitive to the jets originated from sbottom LSP decays, since their decays are 2-body processes to 2 nearly massless particles which will carry an energy $\sim m_{\tilde{b}}/2$, much larger than the 50 GeV threshold. 

We check those points comparing the prediction of the signal 
cross section, calculated as $\sigma(pp\to\tilde{b}\tilde{b}^*)\times BR(\tilde{b}\to b \nu)^2$, with the 95\% observed cross section upper limit, corresponding to sbottom pair production decaying to bottom quarks plus massless neutralinos.

\bigskip
\noindent 
Finally, is worth noting that there are LHC searches looking for events with displaced leptons that can be sensitive to the leptons produced in the decay of a long-lived sbottom LSP. That is the case of the ATLAS search for displaced leptons~\cite{ATLAS:2020wjh}, whose main discriminant is the presence of leptonic tracks with an impact parameter greater than 3 mm. There are other LHC searches that look for displaced signals including leptons, but they target topologies which do not match the decays of the sbottom LSP.
We have tested the sensitivity of the search for displaced leptons trough a recasting strategy similarly to the case of displaced vertices and we found that no additional point is excluded.

\section{Strategy for the scanning}
\label{strategy}

In this section, we describe the methodology that we have employed to search for points of our
parameter space that are compatible with the current experimental data on neutrino and Higgs physics, as well as ensuring that the sbottom is the LSP.
In addition, we have demanded the compatibility with some flavor observables,
such as $B$ and $\mu$ decays.
To this end, we have performed scans on the parameter space of the model, with the input parameters optimally chosen.

\subsection{Experimental constraints}

\label{sec:constr}

All experimental constraints (except the LHC searches which are discussed in the previous section) are taken into account as follows:

\begin{itemize}

\item Neutrino observables\\
We have imposed the results for normal ordering from Ref.~\cite{Esteban:2018azc}, selecting points from the scan that lie within $\pm 3 \sigma$ of all neutrino observables. On the viable obtained points we have imposed the cosmological upper
bound on the sum of the masses of the light active neutrinos given
by $\sum m_{\nu_i} < 0.12$ eV~\cite{Aghanim:2018eyx}.

\item Higgs observables\\
The Higgs sector of the $\mn$ is extended with respect to the (N)MSSM.
For constraining the predictions in that sector of the model, we have interfaced 
{\tt HiggsBounds} {{v}}5.10.2~{\cite{Bechtle:2008jh,Bechtle:2011sb,Bechtle:2013wla,Bechtle:2015pma,Bechtle:2020pkv, Bahl:2021yhk}} with {\tt Multinest}, using a 
conservative $\pm 3 \gev$ theoretical uncertainty on the SM-like Higgs boson in the $\mn$ as obtained with {\tt SPheno}. 
Also, in order to address whether a given Higgs scalar of the $\mn$ 
is in agreement with the signal observed by ATLAS and CMS, we have interfaced
{\tt HiggsSignals} {{v}}2.6.2~{\cite{Bechtle:2013xfa,Bechtle:2020uwn}} with {\tt Multinest}.
Our requirement is that the $p$-value reported by {\tt HiggsSignals} be larger than 2\%, which is equivalent to impose $\chi^2 < 159$ for the 111 relevant degrees of freedom taken into account in our numerical calculation. It is worth noting here that {\tt HiggsTools}~\cite{Bahl:2022igd} was released a year ago, including exotic final states or scalar searches that do not explicitly target Higgs bosons. The inclusion of these computations is not expected to change our current results, thus the implementation of {\tt HiggsTools} is left for future works.

\item $B$~decays\\
$b \to s \gamma$ occurs in the SM at leading order through loop diagrams.
We have constrained the effects of new physics on the rate of this 
process using the average {experimental value of BR$(b \to s \gamma)$} $= (3.55 \pm 0.24) \times 10^{-4}$ provided in Ref.~\cite{Amhis:2012bh}. 
Similarly to the previous process, $B_s \to \mu^+\mu^-$ and  $B_d \to \mu^+\mu^-$ occur radiatively. We have used the combined results of LHCb and CMS~\cite{CMSandLHCbCollaborations:2013pla}, 
$ \text{BR} (B_s \to \mu^+ \mu^-) = (2.9 \pm 0.7) \times 10^{-9}$ and
$ \text{BR} (B_d \to \mu^+ \mu^-) = (3.6 \pm 1.6) \times 10^{-10}$. 
We put $\pm 3\sigma$ cuts from $b \to s \gamma$, $B_s \to \mu^+\mu^-$ and $B_d \to \mu^+\mu^-$, {as obtained with {\tt SPheno}}. {We have also checked that the values obtained are compatible with the $\pm 3 \sigma$ of the recent results from the LHCb collaboration \cite{Santimaria:2021}. 
}

\item $\mu \to e \gamma$ and $\mu \to e e e$\\
We have also included in our analysis the constraints {from 
BR$(\mu \to e\gamma) < 4.2\times 10^{-13}$~\cite{TheMEG:2016wtm}}
and BR$(\mu \to eee) < 1.0 \times 10^{-12}$~\cite{Bellgardt:1987du}, {as obtained with {\tt SPheno}}.

\item Chargino mass bound\\
Charginos have been searched at LEP with the result of a lower limit on the lightest chargino mass of 103.5 GeV in RPC MSSM, assuming universal gaugino and sfermion masses at the GUT scale and electron sneutrino mass larger than 300 GeV~\cite{wg1}. This limit is affected if the mass difference between chargino and neutralino is small, and the lower bound turns out to be in this case 92 GeV~\cite{wg2}. LHC limits  can be stronger but for very specific mass relations~\cite{ATLAS:2019lff,CMS:2018xqw,ATLAS:2017qwn,CMS:2018yan}.
Although in our framework there is RPV and therefore these constraints do not apply automatically, we typically choose in our analyses of the $\mn$   a conservative limit of $m_{\widetilde \chi^\pm_1} > 92 \gev$. However, since in this work we are analysing the sbottom as the LSP, the chargino mass is always well above the mentioned bound.

\item Electroweak precision measurements\\
There have been recently several improvements in EW measurements such as $M_W$, $g-2$, $S, T, U$, etc. (see e.g. Refs.~\cite{Heinemeyer:2006px,Chakraborti:2022vds,Cho:2011rk}). Thus the confrontation of the theory predictions and experimental results might be timely for SUSY models. However, in our framework electroweak precision measurements are not given significant contributions. This is because the SUSY mass spectrum turns out to be above 1.1 TeV, where the latter value is the lower bound that we obtain in Section~\ref{sec:results} for the mass of the sbottom LSP.

\end{itemize}

\subsection{Parameter analysis}
\label{sec:parameter}

\begin{table}[t!]
\begin{center}
\begin{tabular}{|c|c|c|c|c|c|c|c|c|}
\hline
 $\lambda$  & 0.15 &0.20 & 0.25 & 0.30 & 0.35 &  0.40 & 0.45 & 0.50 \\
 \hline
$\tan\beta$ & 9.5 & 7.5 & 4.5 & 3.0 & 2.8 & 2.4 & 2.0 & 1.6 \\
 \hline
\end{tabular}
\end{center}
\caption{
Pair of low-energy values of the input parameters $\lambda$ and $\tan\beta$ determining the eight scans carried out. 
For all the cases, the input parameters $T_{u_3}$, $m_{\widetilde d_{3R}}$ are varied in the ranges shown in Eqs.~(\ref{tu}),~(\ref{sbottommass}), {and $v_{iL}$, $Y_{\nu_i}$ in the ranges shown in Table \ref{tab:neutrino}}.
}
\label{kap06}
\end{table}

\begin{table}[t!]
\begin{center}
\begin{tabular}{|c|}
\hline
$\kappa$ = 0.6\\
$-T_{\kappa}$ = 1000\\
$T_{\lambda}$ = 1500\\
$v_R$ = 3600 \\
$m_{{\widetilde{e}}_{1,2,3R}}=  m_{{\widetilde{d}}_{1,2R}} = m_{{\widetilde{Q}}_{1,2, 3L}} = m_{{\widetilde{u}}_{1,2, 3R}}$ = 2000 \\
$T_{d_{1,2}} = T_{e_{1,2}} = T_{u_{1,2}}$ = 0  \\
$T_{d_3}$ = 100,\ $T_{e_3}$ = 40\\
$-T_{\nu_{1,2,3}}$ = 0.01 \\
$M_1$ = 2400,\ $M_2$ = 2000,\ $M_3$ = 2700 \\
\hline 
\end{tabular}
\end{center}
\caption{Low-energy values of the input parameters that are fixed in the eight scans of Table~\ref{kap06}, with the VEVs $v_R$
and the soft SUSY-breaking parameters given in GeV.}
\label{fixed}
\end{table}

\begin{table}[t!]
    \centering
    \begin{tabular}{|c|}
    \hline
        $v_{1L} \in (6.3\times10^{-5}, 3.1 \times 10^{-4}
        )$ \\
        $v_{2L} \in (1.2 \times 10^{-4}, 7.9\times 10^{-4})$ \\
        $v_{3L} \in (2.5 \times 10^{-4},  1.0 \times 10^{-3})$ \\
        $Y_{\nu_{1}} \in (3.1 \times 10^{-7},  1.0 \times 10^{-6})$ \\
        $Y_{\nu_{2}} \in (1.2 \times 10^{-6},  6.3 \times 10^{-6})$ \\
        $Y_{\nu_{3}} \in (1.5 \times 10^{-9}, 6.3\times10^{-8})$ \\
        \hline
    \end{tabular}
    \caption{{Range of low-energy values of the input parameters related to neutrino physics that are varied in the eight scans of Table~\ref{kap06}, with the VEVs $v_{iL}$ given in GeV.}}
    \label{tab:neutrino}
\end{table}

The parameters $\lambda$ and $\tan\beta$ are crucial for our analysis. First, they contribute to reproduce Higgs data, as discussed in Section~\ref{sec:higgs}. Second, they determine the values of the sbottom decay widths, which depend on the higgsino mass parameter $\mu$ and the bottom Yukawa coupling $Y_b$~(\ref{gamma-bnu}).
Note in this sense that $\lambda$ contributes to $\mu$ (see Eq.~(\ref{mu2})), and that  $Y_b$ increases with $\tan\beta$. 
As it is shown in
Table~\ref{kap06}, 
we chose a range of moderate/large values of $\lambda \in (0.15, 0.50)$, thus we are in a similar situation as in the NMSSM (see Ref.~\cite{Domingo:2019jsc} and references therein) and small/moderate values of $\tan\beta$, $|T_{u_{3}}|$, and soft stop masses are necessary to obtain through loop effects the correct SM-like Higgs mass~{\cite{Biekotter:2017xmf, Biekotter:2019gtq, Kpatcha:2019qsz, Biekotter:2020ehh}}. In particular, the corresponding values of $\tan\beta$ are also shown in Table~\ref{kap06}, 
it is sufficient for our analysis to fix 
$m_{{\widetilde{Q}}_{3L}}$ and $m_{{\widetilde{u}}_{3R}}$ to a reasonable value of
2000 GeV, as can be seen in
Table~\ref{fixed}, and finally for all the cases we scanned over the low-energy values of $T_{u_3}$ in the range:
\bea
 -T_{u_3} = 900-4000\ \text{GeV}.
 \label{tu}
\eea
{It is worth noting here that the entire mass spectrum has been obtained using the full one-loop numerical computation implemented in {\tt SPheno}.

In Table~\ref{fixed}, we also show the low-energy values of other input parameters.
Reproducing Higgs data requires suitable additional parameters such as
$\kappa$, 
$v_R$, $T_\kappa$, $T_\lambda$
(see Eq.~(\ref{freeparameterss})). Thus, we fixed to appropriate values 
$T_\lambda$, which is relevant for obtaining the correct values of the off-diagonal terms of the mass matrix mixing the right sneutrinos with Higgses, and
$\kappa$, $T_\kappa$, $v_R$ which basically control the right sneutrino sector.
To ensure that chargino is heavier than sbottom, the lower value of $\lambda$ forces us to choose a large value for $v_R$ in order to obtain a large enough value of $\mu$ (see Eq.~(\ref{mu2})). 
The parameters $\ka$ and $T_{\kappa}$ are crucial to determine the mass scale of the right sneutrinos.
We choose the value of $-T_{\kappa}$ to have heavy pseudoscalar right sneutrinos,
and therefore 
the value of $\kappa$ has to be large enough in order to avoid 
too light (even tachyonic) scalar right sneutrinos. 
Working with the values of $\lambda$ of Table~\ref{kap06}, we can keep perturbativity up to an intermediate scale of new physics around $10^{11}$~GeV, as discussed in detail in
Ref.~\cite{Kpatcha:2019qsz}.

The values of other parameters shown in Table~\ref{fixed} concern slepton, squark and  gluino masses, as well as quark and lepton trilinear parameters, which are not specially relevant for our analysis. 
The values chosen for the latter are reasonable within the supergravity framework, where
the trilinear parameters are proportional to the corresponding Yukawa couplings. 
Concerning neutrino physics, as discussed in Section~\ref{neutrino} the most crucial parameters~(\ref{freeparameters}) are basically decoupled from those controlling Higgs physics~(\ref{freeparameterss}).
Thus, for the concrete values of $\lambda$, $\kappa$, $\tan\beta$, $v_R$, etc.,
chosen to reproduce Higgs data, there is still enough freedom to reproduce in addition neutrino data by playing with {appropriate values
of $M_1$, $M_2$ and $Y_{\nu_i}$, $v_{iL}$, as shown in the last row of Table~\ref{fixed}, and in Table \ref{tab:neutrino}.}

Finally, the soft mass of the right sbottom, $m_{\widetilde d_{3R}}$,
is obviously a crucial parameter in our analysis, since it controls the physical sbottom mass, as discussed in Section~\ref{sbottom}.
Thus, for obtaining a right sbottom LSP we scanned this parameter in the low-energy range:
\bea
 m_{\widetilde d_{3R}} = 200-2000\ \text{GeV}.
 \label{sbottommass}
\eea

Summarizing, we performed eight scans over the 8 parameters {$m_{\widetilde d_{3R}}$, $T_{u_3}$, $v_{iL}$ and $Y_{\nu_i}$ corresponding to the pair of values ($\lambda$, $\tan\beta$) shown in Table~\ref{kap06}}.

\section{Results}
\label{sec:results}

Following the methods described in the previous sections, in order
to find regions consistent with experimental observations we performed scans of the parameter space, and our results are presented here.
To carry this analysis out, we selected first points from the scans that lie within $\pm 3\sigma$ of all neutrino physics observables \cite{Esteban:2018azc}.
Second, we put $\pm 3\sigma$ cuts from $b \to s \gamma$, $B_s \to \mu^+\mu^-$ and $B_d \to \mu^+\mu^-$ 
and require the points to satisfy also the upper limits of $\mu \to e \gamma$ and $\mu \to eee$. 
In the third step, we imposed that Higgs physics is realized.
In particular, we require that the p-value reported by {\tt HiggsSignals} be larger
than 2\%.
Also, since we are interested in the right sbottom as LSP, of the allowed points we selected those satisfying this condition.

We show in Fig.~\ref{fig:ctau} the proper decay length of the right sbottom LSP 
for the points of the parameter space studied fulfilling the above experimental constraints.
As expected, for a fixed value of $\lambda$ the decay length 
increases with decreasing sbottom mass.
On the other hand, the decay length depends strongly on $\lambda$ (and tan$\beta$).
In particular, for a fixed sbottom mass $c\tau$ increases with increasing $\lambda$. This is because $\lambda$ contributes to $\mu$ and therefore the total decay width (see Eq.~(\ref{gamma-bnu})) decreases with increasing $\lambda$, 
as discussed in Section~\ref{sec:parameter}.
This dependence becomes relevant when applying the LHC constraints discussed in Section~\ref{sec:lhc}.
For $\lambda = 0.15$, $0.20$ all points of our scan have prompt 
decays since $c \tau < 1$~mm,  whereas for 
$\lambda = 0.25-0.4$ 
there are also points with displaced decays, depending on the value of $m_{\widetilde b_{R}}$.
As
shown in the upper plot of the figure,
all points with 
$\lambda = 0.45$, $0.50$ have displaced decays.

As can be seen from Fig.~\ref{fig:ctau}, for the cases $\lambda= 0.25- 0.50$ the maximum value of the sbottom mass as LSP is $\sim 1900$ GeV for the chosen scan range in Eq.~(\ref{sbottommass}). The slight shift in the upper bound is because for masses close to 2000 GeV the impact of $\tan\beta$ in the stop mass is relevant. In particular, when increasing $\lambda$ smaller values of $\tan\beta$ are necessary to reproduce the Higgs mass implying in turn smaller values for the stop mass, resulting at the end of the day in a stop LSP. On the other hand, for $\lambda=0.15, 0.20$, 
the maximum values of the sbottom LSP masses are $\sim 1200$, $1500$ GeV, respectively.
This is because for these masses and values of $\lambda$ and $v_R$
the $\mu$ parameter becomes sufficiently small as to give rise to neutral higgsino LSPs. For $\lambda<0.15$ the neutral higgsino is the LSP unless the sbottom is very light, and, as a consequence experimentally excluded.
It is true that choosing larger values of $v_R$ would allow larger higgsino masses, modifying this lower bound for $\lambda$. Nevertheless, given the contribution of both parameters to $\mu$, this would be equivalent to increase $\lambda$, and, as will be clear from the discussion below, the relevant lower bounds for the sbottom LSP mass found would not change.

\begin{figure}[t!]
\centering
        
    \begin{subfigure}[b]{0.7\textwidth}
  \includegraphics[width=\textwidth]{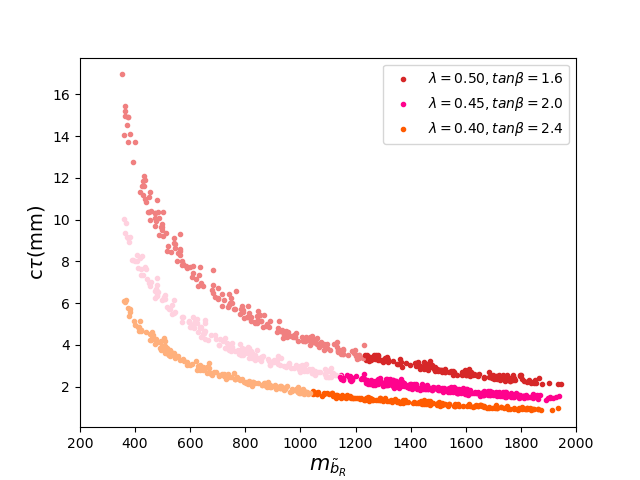}
  \label{fig:sub2}
\end{subfigure}

    \begin{subfigure}[b]{0.7\textwidth}
 \includegraphics[width=\textwidth]{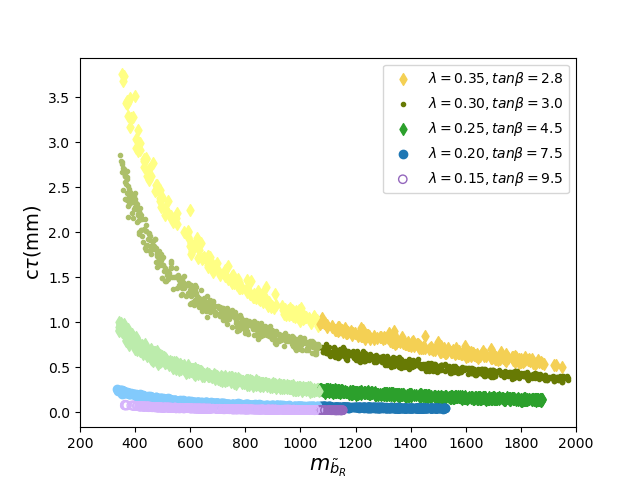}
 
  \label{fig:sub1}
\end{subfigure}
 \caption{Proper decay length $c\tau_{\widetilde b_{R}}$ 
versus the right sbottom mass $m_{\widetilde b_{R}}  \; [ \text{GeV} ]$,
corresponding to the scans discussed in Section~\ref{sec:parameter} 
with $\lambda=0.15, 0.20, 0.25, 0.30, 0.35$ (lower plot) and $\lambda=0.40, 0.45, 0.50$ (upper plot).
    All points fulfill the experimental constraints discussed in Section~\ref{sec:constr}. 
(Light) Dark points (do not) fulfill the LHC constraints. 
}
\label{fig:ctau}
\end{figure}

In Fig.~\ref{fig:branging ratios}, the branching ratios (BRs) of both decay modes corresponding to each $\lambda$ are shown, i.e. sbottom LSP decaying to a top and leptons (lower plots) and decaying to a bottom and neutrinos (upper plots). 
First, we see that the BR of $\widetilde{b}_R$ to leptons is smaller than the one to neutrinos for sbottom masses close to the top mass. 
As discussed in Section.~\ref{subsec:Decaymodes}, this is an obvious consequence of Eq.~(\ref{gamma-bnu}) for the partial decay widths.
Second, for a fixed $m_{\widetilde b_{R}}$ the BR of $\widetilde{b}_R$ to leptons (neutrinos) decrease (increase) with increasing (decreasing) $\lambda$. This is because the decay width to leptons is inversely proportional to $\lambda$, while the total decay width is slightly dominated by the term proportional to $U^V_{i4}$ in the decay width to neutrinos. 
The latter occurs because even though neutrino-higgsino mixing is slightly bigger than the neutrino-bino one, the couplings multiplying them make $\sqrt{2}g'U^V_{i4}/3\gtrsim Y_b U^V_{i6}$. In addition, this term makes that the decay width to neutrinos slightly dominates with respect to the one to leptons. 
As can be seen from Fig.~\ref{fig:branging ratios}, for each decay channel the variation of BRs with $\lambda$ is less than 10 $\%$.

\begin{figure}[t!]
\centering

    \begin{subfigure}[b]{1.1\textwidth}
  \includegraphics[width=\textwidth]{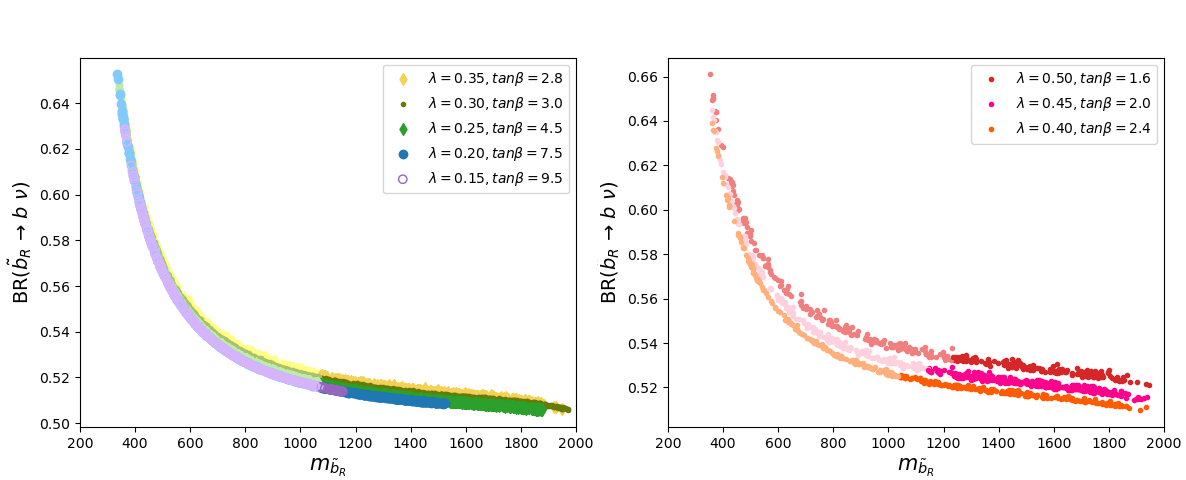}
  \label{fig:sub2}
\end{subfigure}

    \begin{subfigure}[b]{1.1\textwidth}
 \includegraphics[width=\textwidth]{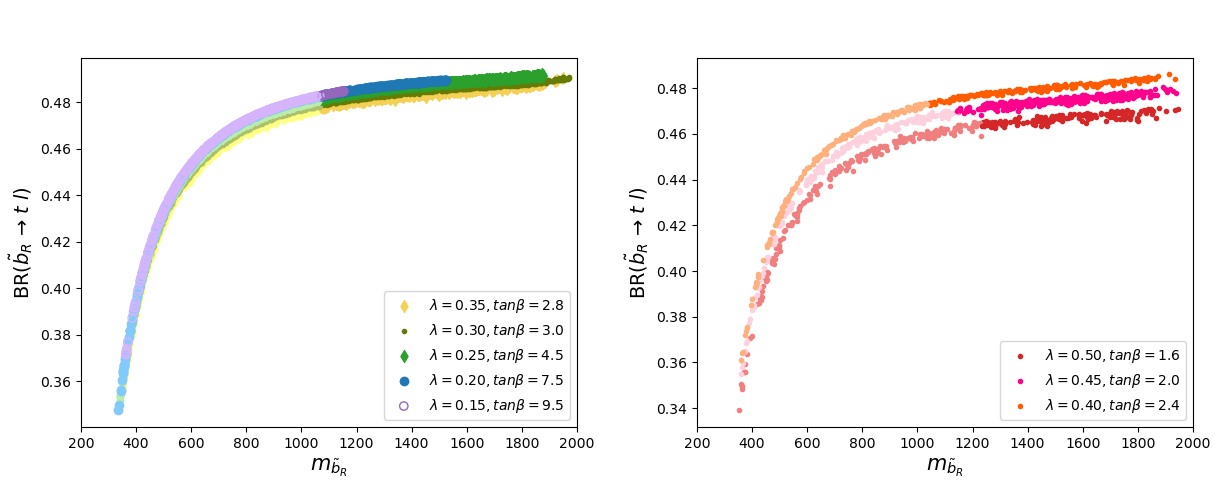}
  \label{fig:sub1}
\end{subfigure}
\caption{Points of Fig.~\ref{fig:ctau} but showing for them the sum of branching ratios of right sbottom LSP decaying to $t \ell$ (lower plots) 
and $b \nu$ (upper plots) versus the right sbottom mass $m_{\widetilde b_{R}} \; [ \text{GeV} ]$.
(Light) Dark points (do not) fulfill the LHC constraints.
}
\label{fig:branging ratios}
\end{figure}

In these figures, for the points of our scans with displaced decay lengths we applied the LHC constraints discussed in \textit{Case (ii)} 
 of Sec.~\ref{sec:lhc}. {As we can see from Fig.~\ref{fig:ctau}, $c\tau$ is well below 300 mm for all points, therefore \textit{Case (i)} is not used to constrain any of them.} For the points of our scans with prompt decay lengths we applied the constraints discussed in \textit{Case (iii)}. 
As a result of our analysis, points with dark (light) colours in the figures are allowed (forbidden) by LHC data.
In particular, in the lower plot of 
Fig.~\ref{fig:ctau} allowed (dark) points start to appear for 
$m_{\tilde{b}_R}\approx 1070$~GeV,
corresponding to prompt decay lengths of 0.97~mm for $\lambda=0.35$, 
0.73~mm for $\lambda=0.30$, 0.26~mm for $\lambda=0.25$,  0.06~mm for $\lambda=0.20$, and 0.02~mm for $\lambda=0.15$. 
{In the upper plot,
for $\lambda=0.40$ 
the allowed points start to appear for
$m_{\tilde{b}_R}\approx 1041$~GeV, corresponding to a decay length of 1.7~mm. For $\lambda=0.45$, they appear for
$m_{\tilde{b}_R}\approx 1130$~GeV, with decay length of 2.6~mm. Finally, in the case of $\lambda=0.5$, this happens for 
$m_{\tilde{b}_R}\approx 1235$~GeV, corresponding to a decay length of 3.5~mm.
It is worth noting that for $\lambda=0.40$ the decay length becomes smaller than 1~mm for masses $m_{\tilde{b}_R}\approx 1685$~GeV, so for the analysis 
we combined the constraints of \textit{Cases (ii)} 
and~\textit{(iii)}}.

Let us finally remark that the use of other type of solutions for neutrino physics different from the one presented in Eq.~(\ref{neutrinomassess}), would not modify the results obtained. {This can be understood from the summation over leptons present in Eqs.~\ref{gamma-bnu} and \ref{--sneutrino-decay-width-2nus2}, since for the most restrictive searches, for instance \cite{ATLAS:2017tny} \cite{ATLAS:2020wjh}, the results are independent of the lepton family or integrate over it.
}

\section{Conclusions}
\label{sec:conclusions}
We analyzed 
the signals expected at the LHC for a right sbottom LSP
in the framework of the $\mn$, imposing on the parameter space the experimental constraints on neutrino and Higgs physics, as well as flavour observables such as $B$ and $\mu$ decays. The sbottoms are pair produced and have two different decay channels producing a lepton and a top quark, or a neutrino and a bottom quark. We studied these channels and the corresponding decay length for different representative values of the trilinear coupling $\lambda$ between right sneutrinos and Higgses, comparing
the predictions with ATLAS and CMS results~\cite{CMS:2019qjk,ATLAS:2017tny,ATLAS:2021yij,ATLAS:2013lcn}.}
As shown in Fig.~\ref{fig:ctau}, for $\lambda\in (0.15-0.35)$ the allowed points have prompt decays, and we obtained a lower limit on the sbottom mass of about 1070~GeV. On the other hand, for $\lambda\in (0.40-0.50)$ we found that the allowed points
have displaced decays, and a lower limit on the sbottom mass of about 1041 GeV was obtained. The largest value for the decay length found 
is about 3.5~mm.


\begin{acknowledgments}

The work of P.K. and D.L. was supported by the Argentinian CONICET, and they also acknowledge the support through
{PICT~2020-02181}.  
The work of E.K. was supported by the grant "Margarita Salas" for the training of young doctors (CA1/RSUE/2021-00899), co-financed by the Ministry of Universities, the Recovery, Transformation and Resilience Plan, and the Autonomous University of Madrid.
The work of I.L.\ was funded by the Norwegian Financial Mechanism 2014-2021, grant DEC-2019/34/H/ST2/00707. 
The research of C.M. was partially supported by the AEI
through the grants IFT Centro de Excelencia Severo Ochoa No CEX2020-001007-S and PID2021-125331NB-I00, funded by MCIN/AEI/10.13039/501100011033.

\end{acknowledgments}


\appendix
\numberwithin{equation}{section}
\numberwithin{figure}{section}
\numberwithin{table}{section}

\numberwithin{equation}{subsection}
\numberwithin{figure}{subsection}
\numberwithin{table}{subsection}

\section{One Down Squark-two Fermion--Interactions} 
\label{appendix}

In this Appendix we write the relevant interactions for our computation of the decays of the sbottom LSP, 
following {\tt SARAH} notation~\cite{Staub:2013tta}.
In particular, 
now
$a,b=1,2,3$ are family indexes, $i,j,k$ are the indexes for the physical states, and $\alpha,\beta,\gamma=1,2,3$ are $SU(3)_C$ indexes.
The matrices $Z^D, U^d_{L,R}, U^u_{L.R}, U^e_{L,R}$ and $U^V$ diagonalize the mass matrices of down squarks, down quarks, up quarks, charged fermions (leptons, gauginos and higgsinos) and neutral fermions (LH and RH neutrinos, gauginos and higgsinos), respectively.
More details about these matrices can be found in Appendix B of Ref.~\cite{Ghosh:2017yeh}.
Taking all this into account, in the basis of 4--component spinors with the projectors 
$P_{L,R}=(1\mp\gamma_5)/2$, the interactions for the mass eigenstates are as follows.

\subsection{Down squark - up quark - lepton Interaction }

\begin{center} 
\begin{fmffile}{Figures/Diagrams/FeynDia119} 
\fmfframe(20,20)(20,20){ 
\begin{fmfgraph*}(75,75) 
\fmfleft{l1}
\fmfright{r1,r2}
\fmf{fermion}{v1,l1}
\fmf{fermion}{v1,r1}
\fmf{scalar}{r2,v1}
\fmflabel{$\bar{e}_{{i}}$}{l1}
\fmflabel{$\bar{u}_{{j \beta}}$}{r1}
\fmflabel{$\tilde{d}_{{k \gamma}}$}{r2}
\end{fmfgraph*}} 
\end{fmffile} 
\end{center} 

\vspace*{-0.5cm}

\begin{align} 
\nonumber  & i \delta_{\beta \gamma}\Big(U^{e,*}_{L,{i 5}}  \sum_{b=1}^{3}Z^{D,*}_{k b} \sum_{a=1}^{3}U^{u,*}_{R,{j a}} Y_{u,{a b}}\Big) P_L\\ 
  &  -\,i \delta_{\beta \gamma} \Big(g\sum_{a=1}^{3}Z^{D,*}_{k a} U_{L,{j a}}^{u}  U_{R,{i 4}}^{e}  - \sum_{b=1}^{3}\sum_{a=1}^{3}Y^*_{d,{a b}} Z^{D,*}_{k 3 + a}  U_{L,{j b}}^{u}  U_{R,{i 5}}^{e} \Big)P_R.
\label{eq:appen-ql}
  \end{align}

\subsection{Down squark - down quark - neutrino Interaction }

\begin{center} 
\begin{fmffile}{Figures/Diagrams/FeynDia108} 
\fmfframe(20,20)(20,20){ 
\begin{fmfgraph*}(75,75) 
\fmfleft{l1}
\fmfright{r1,r2}
\fmf{fermion}{v1,l1}
\fmf{plain}{r1,v1}
\fmf{scalar}{r2,v1}
\fmflabel{$\bar{d}_{{i \alpha}}$}{l1}
\fmflabel{$\nu_{{j}}$}{r1}
\fmflabel{$\tilde{d}_{{k \gamma}}$}{r2}
\end{fmfgraph*}} 
\end{fmffile} 
\end{center} 

\vspace*{-0.5cm}
  \begin{align} 
\nonumber   &-\frac{i}{3} \delta_{\alpha \gamma} \Big(3 U^{V,*}_{j 6} \sum_{b=1}^{3}Z^{D,*}_{k b} \sum_{a=1}^{3}U^{d,*}_{R,{i a}} Y_{d,{a b}}    + \sqrt{2} g' U^{V,*}_{j 4} \sum_{a=1}^{3}Z^{D,*}_{k 3 + a} U^{d,*}_{R,{i a}}  \Big)P_L\\ 
    &  \,-\frac{i}{6} \delta_{\alpha \gamma} \Big(6 \sum_{b=1}^{3}\sum_{a=1}^{3}Y^*_{d,{a b}} Z^{D,*}_{k 3 + a}  U_{L,{i b}}^{d}  U_{{j 6}}^{V}  + \sqrt{2} \sum_{a=1}^{3}Z^{D,*}_{k a} U_{L,{i a}}^{d}  \Big(-3 U_{{j 5}}^{V}  + g' U_{{j 4}}^{V} \Big)\Big)P_R.
  \label{eq:appen-qnu}
  \end{align}


\clearpage
\bibliographystyle{utphys}
\bibliography{munussmbib-completo_v6}


\end{document}